\documentclass[a4paper,conference]{IEEEtran}
\usepackage[T1]{fontenc}
\usepackage[latin9]{inputenc}
\usepackage{units}
\usepackage{algorithm2e}
\usepackage{amsmath}
\usepackage{amsthm}
\usepackage{graphicx}

\makeatletter

\pdfpageheight\paperheight
\pdfpagewidth\paperwidth

\providecommand{\tabularnewline}{\\}

\theoremstyle{plain}
\newtheorem{thm}{\protect\theoremname}
\theoremstyle{definition}
\newtheorem{defn}[thm]{\protect\definitionname}

\usepackage{cite}
\usepackage{amsmath,amssymb,amsfonts}
\usepackage{algorithmic}
\usepackage{graphicx}
\usepackage{textcomp}
\usepackage{xcolor}
\def\BibTeX{{\rm B\kern-.05em{\sc i\kern-.025em b}\kern-.08em
    T\kern-.1667em\lower.7ex\hbox{E}\kern-.125emX}}

\LinesNumbered
\SetKwRepeat{Do}{do}{while}%

\let\oldnl\nl
\newcommand{\nonl}{\renewcommand{\nl}{\let\nl\oldnl}}

\@ifundefined{showcaptionsetup}{}{%
 \PassOptionsToPackage{caption=false}{subfig}}
\usepackage{subfig}
\makeatother

\providecommand{\definitionname}{Definition}
\providecommand{\theoremname}{Theorem}

\begin{document}
\title{Joint Compressed Sensing and Manipulation of Wireless Emissions with
Intelligent Surfaces}
\author{Christos Liaskos$^{1}$, Ageliki Tsioliaridou$^{1}$, Alexandros Pitilakis$^{2}$,
George Pirialakos$^{2}$, Odysseas Tsilipakos$^{1}$,\\
Anna Tasolamprou$^{1}$, Nikolaos Kantartzis$^{2}$, Sotiris Ioannidis$^{1}$,
Maria Kafesaki$^{1}$, \\
Andreas Pitsillides$^{3}$ and Ian Akyildiz$^{3}$\\
$^{1}${\small{}Foundation for Research and Technology - Hellas (FORTH)}\\
{\small{}Emails: \{cliaskos,atsiolia\}@ics.forth.gr, otsilipakos@iesl.forth.gr,
atasolam@iesl.forth.gr, sotiris@ics.forth.gr, kafesaki@iesl.forth.gr}\\
$^{2}${\small{}Aristotle University of Thessaloniki - Hellas (AUTH)}\\
{\small{}Emails: alexpiti@ece.auth.gr, pyrialak@auth.gr, kant@auth.gr}\\
$^{3}${\small{}University of Cyprus (UCY)}\\
{\small{}Emails: Andreas.Pitsillides@ucy.ac.cy, ian@ece.gatech.edu}}

\maketitle
\textbf{\small{}Abstract\textemdash Programmable, intelligent surfaces
can manipulate electromagnetic waves impinging upon them, producing
arbitrarily shaped reflection, refraction and diffraction, to the
benefit of wireless users. Moreover, in their recent form of HyperSurfaces,
they have acquired inter-networking capabilities, enabling the Internet
of Material Properties with immense potential in wireless communications.
However, as with any system with inputs and outputs, accurate sensing
of the impinging wave attributes is imperative for programming HyperSurfaces
to obtain a required response. Related solutions include field nano-sensors
embedded within HyperSurfaces to perform minute measurements over
the area of the HyperSurface, as well as external sensing systems.
The present work proposes a sensing system that can operate without
such additional hardware. The novel scheme programs the HyperSurface
to perform compressed sensing of the impinging wave via simple one-antenna
power measurements. The HyperSurface can jointly be programmed for
both wave sensing and wave manipulation duties at the same time. Evaluation
via simulations validates the concept and highlight its promising
potential.}{\small\par}

\section{Introduction\label{SECINTRO}}

Electromagnetic (EM) wave propagation along a wireless channel exhibits
fundamental and well-studied phenomena that hinder wireless communications:
Path loss, multi-path fading and Doppler shift are presently unsurmountable,
degenerative factors that cannot be controlled. Thus, communication
system designers seek to adapt to them as best as possible, much like
surviving a tropical storm. Hence in order to compensate for this
unpredictable wireless channel behavior, exacerbated by other users
and uncontrollable environmental factors, they act as the devices
on the edge. Notice that the hunt for higher data rates in the upcoming
5th Generation of wireless communications (5G) pushes for very high
communication frequencies, e.g., at 60 GHz, where the described effects
become extremely acute, and especially at the large scales imposed
by IoT~\cite{akyildiz20165g}.

\begin{figure}[!t]
\begin{centering}
\includegraphics[width=1\columnwidth]{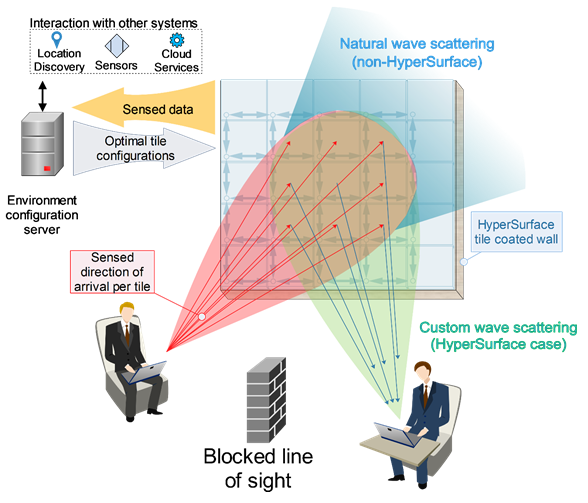}
\par\end{centering}
\caption{\label{fig:OverviewPWE}The proposed approach involving HyperSurface
tile-coated environmental objects. The wireless propagation is tailored
to the needs of the communication link under optimization. Unnatural
propagation, such as lens-like focus and negative reflection angles
can be provided. }
\end{figure}

\begin{figure}
\begin{centering}
\includegraphics[width=1\columnwidth]{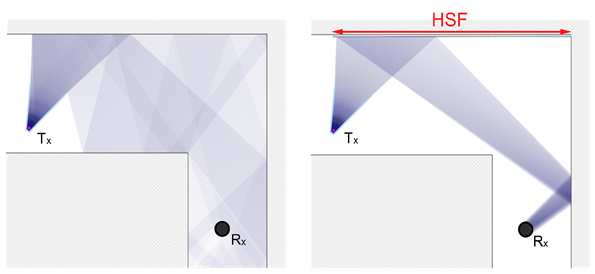}
\par\end{centering}
\caption{\label{fig:focusExample}HyperSurfaces (HSF) can counter free space
path loss by acting as a lens for impinging waves. A ray tracing-derived
showcase with a single transmitter (Tx) / receiver (Rx) pair is shown.
Darker colors denote higher signal power levels.}
\end{figure}

A recently proposed approach for wireless communications is concept
of programmable wireless environments~\cite{CACM.2018,COMMAG.2018}.
This novel approach can readily combat path loss, multi-path fading
and the Doppler shift; an example is shown in Fig.~\ref{fig:OverviewPWE}.
HyperSurface (HSF) tiles, a class of software adaptive metasurfaces,
briefly described later, coat a wall and sense the direction of EM
waves impinging upon them~\cite{wowmom.2018}. The tiles are networked
to one another and to the external world. The sensed data are relayed
to an environmental configuration server that decides upon the optimal
EM behavior to be deployed within the environment, and sends the corresponding
configuration directives to each tile. For example, path loss can
be readily negated as shown in Fig.~\ref{fig:focusExample}. Instead
of ever-dissipating in the environment, the data-carrying wave is
focused in a lens-like manner to the intended target. Moreover, the
focused waves can bounce across several tile-coated objects at\textendash previously
impossible\textendash angles, reaching remote, non-line of sight areas.
HyperSurfaces epitomize the granularity in controlling electromagnetic
waves. This allows for minute control over the echoes reaching an
intended receiver, mitigating the path-loss effect. Moreover, the
lens focal point can be altered in real-time to match the velocity
of a moving target, battling Doppler effects~\cite{tsilipakos2018pairing}.

The derived practical benefits are highly promising. As shown in Fig.~\ref{fig:focusExample},
path loss mitigation results in less power scattering, and increased
received power level. This readily allows for lower-power transmissions,
which favor the battery lifetime of IoT devices. Moreover, the decreased
scattering reduces cross-device interference, allowing an increased
number of mobile users to co-exist in the same space, without degrading
their performance. Additionally, the traveling wave reaches the receiver
via well-defined paths rather than via multiple echoes, allowing for
increased data transmission rates and high-quality coverage even at
previously \textquotedblleft hidden\textquotedblright{} areas. From
another aspect, this separation of user devices can target increased
privacy. Waves carrying sensitive data can be tuned to avoid all other
devices apart from the intended recipient, hindering eavesdropping~\cite{Liaskos2019ADHOC}.
This compliments the security of IoT devices, where hardware restrictions
hinder robust security. These interesting environmental behaviors
and more can be expressed in software in the form of combinable and
reusable modules. Thus, communication system designers and operators
are enabled to easily and jointly optimize the complete data delivery
process, including the wireless environment, supplementing the customizable
wireless device behavior, and furthermore, reducing the complexity
of the device design. This new research direction is essentially \emph{the
Internet of Materials}, which highlights the interconnection of material
properties into smart control loops.

Programming and manipulating the wireless propagation and its effect
requires precise sensing of emitted waves in the first place. This
can be accomplished by employing external systems~\cite{Liaskos2019ADHOC},
such as device localization systems~\cite{shi2018accurate}, and
deduce the nature of their emissions, or by incorporating EM field
sensors within HyperSurfaces~\cite{nanocom.2017}. While valid, these
solutions introduce the complexity of adding new hardware and orchestrating
different systems. The present work contributes a wave sensing system
for use in intelligent environments that does not have such restrictions.
Common signal power level measurements taken by a single antenna/device
are shown to suffice for reconstructing the EM wavefront impinging
on a HyperSurface, and manipulating it accordingly. Moreover, the
same HyperSurface can execute both tasks (i.e., sensing and manipulating
waves) at the same time. The methodology follows the principles of
compressed sensing and RF single-radar imaging~\cite{duarte2008single,wallace2010analysis}.
First, we define a HyperSurface configuration that yields a required
functionality, such as wave steering. Then, we define a series of
additional configurations that can be employed for RF wavefront sensing,
and interleave them via a novel approach.

The remainder of this paper is as follows. Section~\ref{sec:prereqs}
provides the necessary background knowledge and surveys related studies.
Section~\ref{sec:Proposed} presents the proposed scheme. Evaluation
follows in Section~\ref{sec:Evaluation}, and the paper is concluded
in Section~\ref{sec:Conclusion}.

\section{Background and related works\label{sec:prereqs}}

In this Section we briefly describe the operating principles of HSFs,
as well as the concept of compressed sensing, to a level appropriate
for the present work. The reader is also redirected below to related
studies for further information on these topics.

\subsection{HyperSurfaces\label{subsec:metasurf}}

\begin{figure}
\begin{centering}
\includegraphics[width=1\columnwidth]{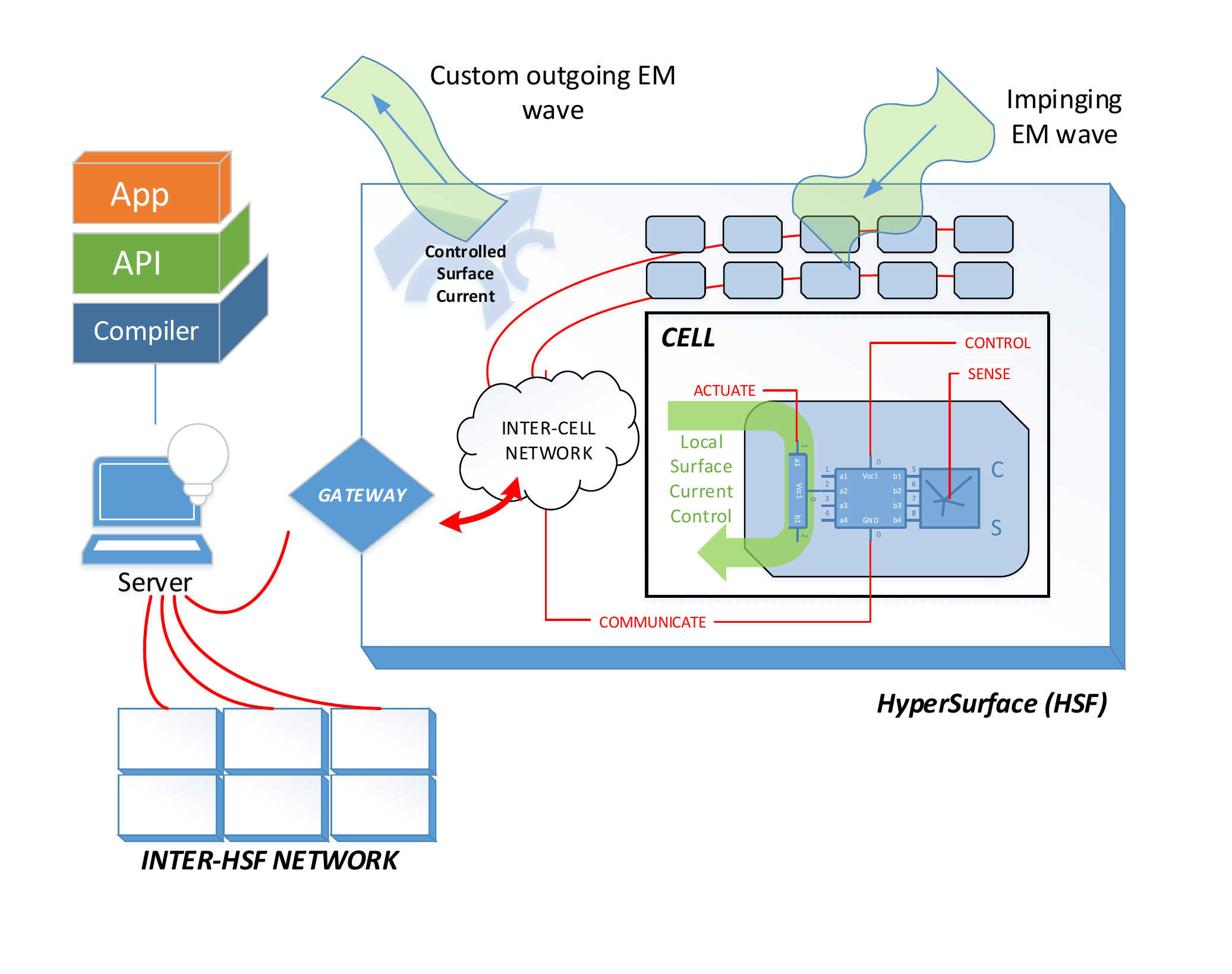}
\par\end{centering}
\caption{\label{fig:HSFcompo}The HyperSurface components, inter-connectivity
and operation principles.}
\end{figure}
The core functionality of HSFs relies on a basic principle in Physics,
which states that the EM emissions from a surface are fully defined
by the distribution of electrical currents over it. The cause that
produces the surface currents are impinging EM waves. Thus, HSFs seek
to control and modify the current distribution over them, in order
to produce a custom EM emission as a response. This outcome can include
any combination of steering, absorbing, polarization and phase alteration
and frequency-selective filtering over the original impinging wave,
even in ways not found in nature~\cite{MSSurveyAllFunctionsAndTypes}.

As shown in Fig.~\ref{fig:HSFcompo}, A HSF comprises a massively
repeated \emph{cell} structure (also known as \emph{meta-atom}), which
includes:
\begin{itemize}
\item passive conductive elements that can be perceived as receiving/transmitting
antennas for the impinging waves,
\item an actuation module, which can regulate the local current flow within
the cell vicinity (e.g., a simple ON/OFF switch),
\item a computation and communication module responsible for controlling
the sensory and actuation tasks within the cell, as well as exchange
data with other cells (inter-cell networking) to perform synergistic
tasks with other cells and communicate with HSF-external entities,
and
\item optionally, a sensor module to detect the attributes of the impinging
wave at the cell vicinity. \emph{Notice that the present work does
not consider such sensors}.
\end{itemize}
Each HSF unit has a gateway that handles its connectivity to the external
word. The gateway participates in the inter-cell network as a peer,
and to the external world via any common protocol (e.g., WiFi or Ethernet).
Its overall role is to:
\begin{itemize}
\item aggregate and transfer sensory data from the HSF to an external controller,
and
\item receive cell actuation commands and diffuse them for propagation within
the inter-cell network.
\end{itemize}
Finally, a regular computer can act as the external entity that gathers
the sensory information from all HSFs within an environment, and subsequently
calculates their configurations that fit a given application scenario.
For instance, programmable wireless environments use multiple HSFs
to customize the wireless propagation for multiple mobile devices,
thus achieving state-of-the-art communication quality~\cite{Liaskos2019ADHOC},
as conceptually shown in Fig.~\ref{fig:OverviewPWE}.

HSFs come with software libraries that facilitate the creation of
applications. This software suite comprises the HSF Application Programming
Interface (API)~\cite{LiaskosAPI}, and the HSF Electromagnetic Compiler~\cite{LiaskosComp}.
The HSF API contains software descriptions of the metasurface electromagnetic
functions and allows the programmer to customize, deploy or retract
them on demand via a programming interface with appropriate callbacks.
The API serves as a strong layer of abstraction. It hides the internal
complexity of the HSF and offers general purpose access to metasurface
functions without requiring knowledge of the underlying hardware and
physics. The EM Compiler handles the translation of the API callbacks
into HSF actuation directives, in an automatic manner, transparently
to the user.
\begin{defn}
The set of actuation element states corresponding to a given EM functionality
(such as wave steering) will be referred to as the HSF \emph{configuration}
for this function.
\end{defn}
\begin{figure}
\begin{centering}
\includegraphics[width=1\columnwidth]{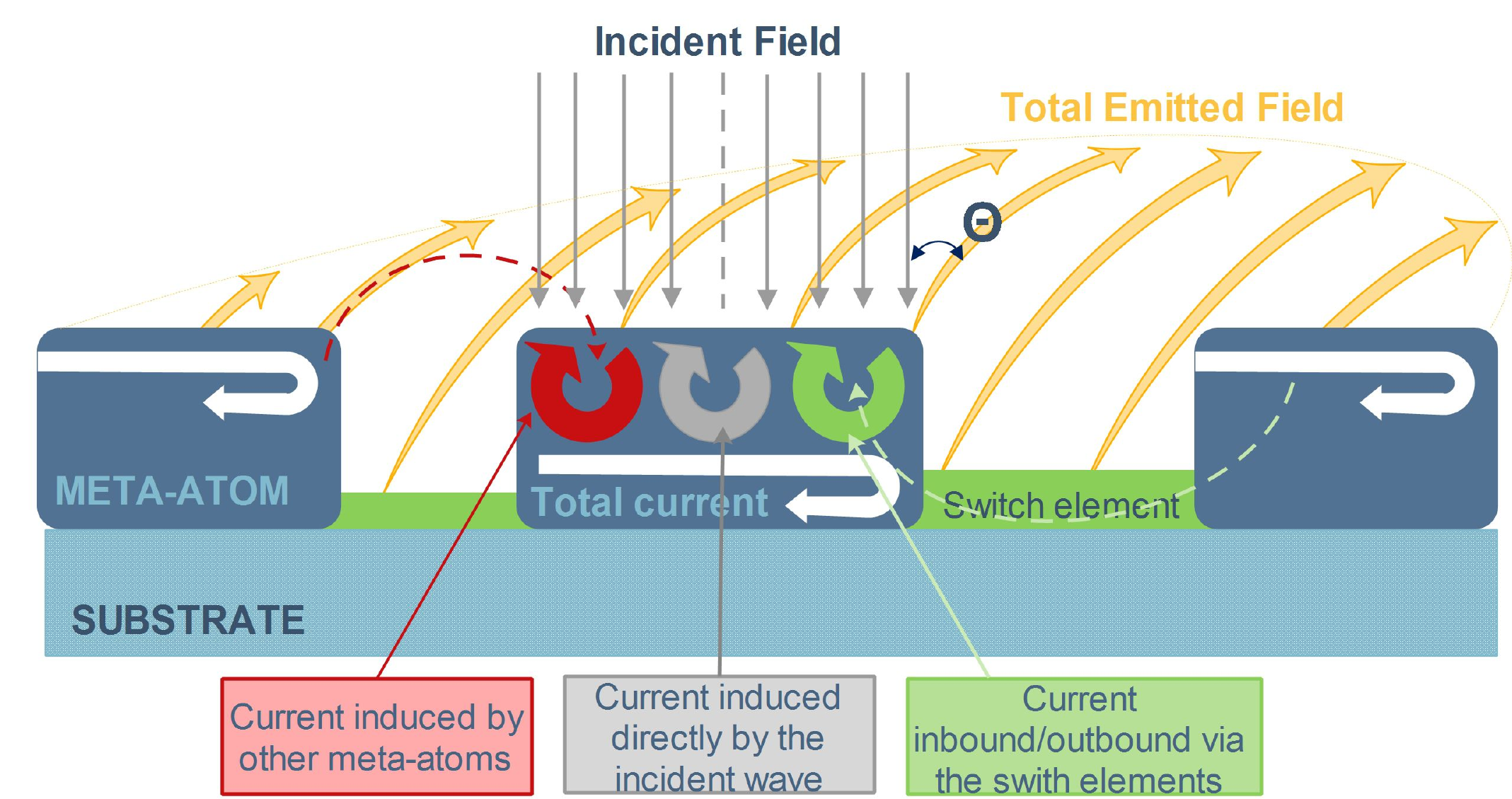}
\par\end{centering}
\caption{\label{fig:MetaSurfCurrents}The HyperSurface components, inter-connectivity
and operation principles.}
\end{figure}
One final note on HSFs and metasurfaces in general, is the strong
EM interactions between the various cells and components. As shown
in Fig.~\ref{fig:MetaSurfCurrents}, an impinging (incident) EM field
creates direct inductive currents to all HSF components. These direct
currents also affect each other via induction, and their values stabilize
after a transient phase (whose duration is negligible). The states
of the switches also affect the current flow distribution, leading
to the total outcome (emitted field).

\subsection{Compressed Sensing \label{subsec:Compressive-Sensing}}

Compressed sensing (CS) is a mathematical tool that can be used to
sample a signal below the Nyquist rate, while preserving its reconstruct-ability,
without significant loss of precision~\cite{donoho2006compressed}.
CS works on sparse signals, i.e., vectors comprising mostly zeros
under some representation.

Let $\boldsymbol{x}$ denote one sparse vector of size $N\times1$.
According to CS, the signal undergoes a sampling process than can
be represented by an under-determined linear system:
\begin{equation}
{\displaystyle \boldsymbol{o}_{K\times1}=A_{K\times N}\cdot\boldsymbol{x}{}_{N\times1}}\label{eq:CSmatrix}
\end{equation}
where~$K<N$. As a rule of a thumb,~$K$ is usually in the range
of~$10\text{\%}-25\text{\%}\cdot N$. The vector~$\boldsymbol{o}$
holds the samples or observations of the original vector~$\boldsymbol{x}$.
The sampling matrix~$A$ needs to be uphold some analytical criteria
described in~\cite{candes2008introduction}.

The reconstruction process involves the solution of an under-determined
linear system, which is naturally possible only by adding additional
restrictions to the solution. In the CS case, the restriction is to
minimize the number of non-zero elements of~$\boldsymbol{x}$. To
this end, a reconstruction process begins with an initial estimate:
\begin{equation}
\boldsymbol{x_{e}^{o}}=A^{T}\cdot\boldsymbol{o}
\end{equation}
and proceeds to iteratively ``punish'' non-zero elements of~$\boldsymbol{x_{e}}$,
with an overall objective to minimize its $L_{1}$ norm~\cite{tsaig2006extensions},
i.e:
\begin{equation}
{\displaystyle L_{1}\left(\boldsymbol{x_{e}}\right)=|x_{1}|+|x_{2}|+\dotsb+|x_{n}|.}
\end{equation}
The $L_{1}$ minimization objective has been shown to lead to robust
and very precise reconstruction outcomes~\cite{donoho2006compressed}.
Existing, free software packages can generate the matrix $A$ and
reconstruct the original vector~$\boldsymbol{x}$ from the observations~$\boldsymbol{o}$~\cite{brad}.
\begin{figure}
\begin{centering}
\includegraphics[width=1\columnwidth]{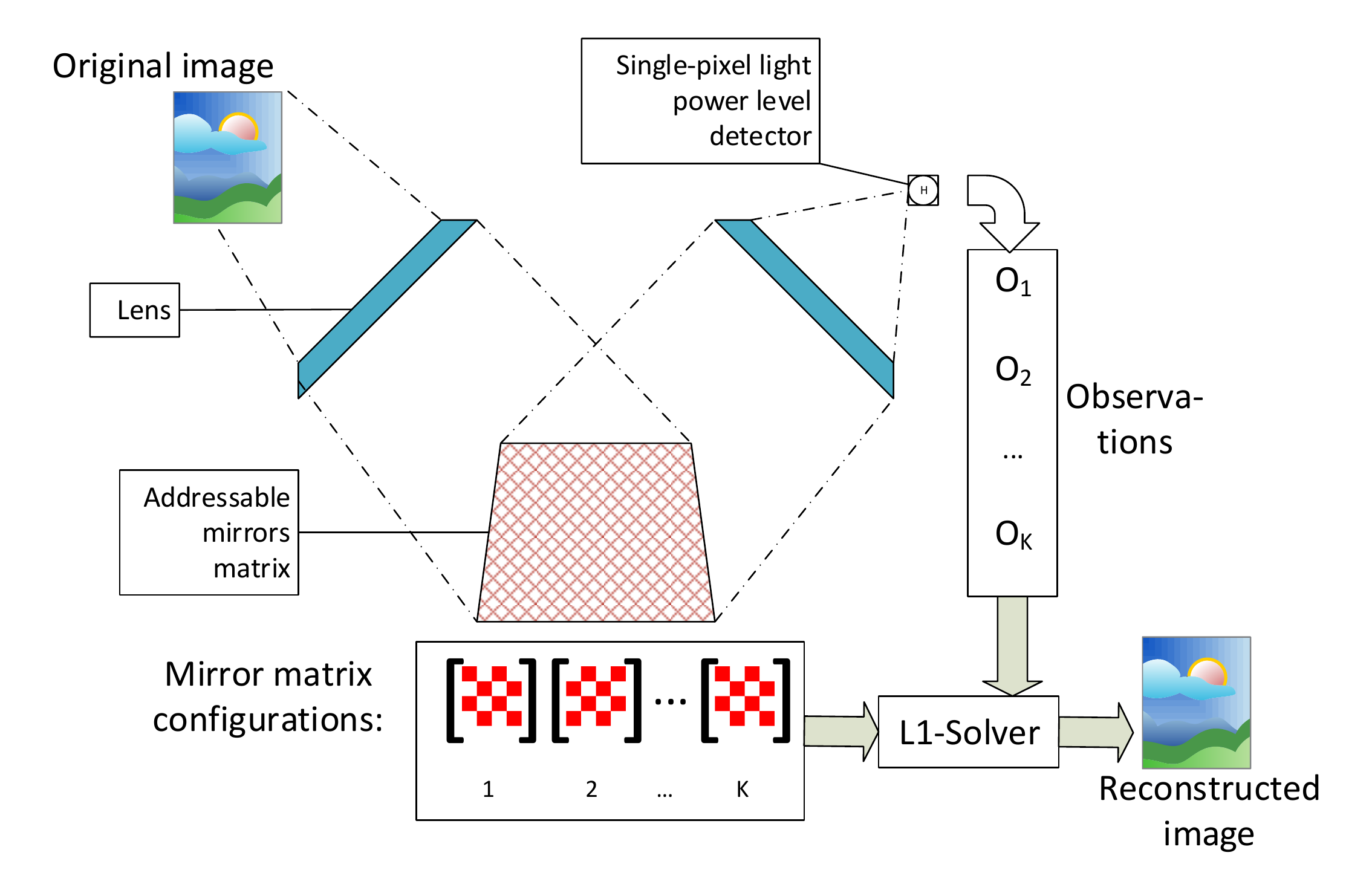}
\par\end{centering}
\caption{\label{fig:onePixelCam}The 1-pixel camera workflow for obtaining
visible light images. There is no interaction between the mirror matrix
elements, and their reflections are completely independent from each
other at the physical layer.}
\end{figure}
A well-known application of CS is the concept of the single pixel
camera~\cite{duarte2008single}, whose composition and operation
principle is shown in Fig.~\ref{fig:onePixelCam}. At the core of
this camera there exists an array of micro-mirrors, whose orientation
can be individually and programmatically tuned (usually by binary
flipping between $\pm15^{o}$ tilt). An image carried by visible light
is sent on this mirror array by means of a simple passive lens. The
aggregate reflection outcome is focused on a single photodiode at
any given time, again via a standard lens. The $\pm15^{o}$ mirror
tilt is sufficient for sending a single ray on or off the photodiode.
By repeating this process several times for different mirror arrangements,
one can realize the sampling matrix of a compressed sensing system.

The same principles have been applied to RF imaging, via a concept
known in this spectrum as single-radar imaging~\cite{wallace2010analysis}.
In the RF case, the imaging process refers to the reconstruction of
the wavefront that reaches a single user antenna. As in the 1-pixel
camera, an actuation device similar to the mirror matrix is required,
to generate the random multiple-mode modulators required for performing
compressed sensing. Programmable metasurfaces have been shown to efficiently
fit this role~\cite{chan2009spatial,watts2012metamaterial,sensale2013terahertz}.
RF imaging with metasurfaces has been successfully used for approximating
the shape of planar metallic objects in space, by emitting waves that
impinge upon these objects and then sensing and reconstructing the
resulting wavefront~\cite{li2016transmission}.

In the present work, we extend the related studies by combining wave
manipulation and wavefront sensing at the same time, using the capabilities
of HSFs. Thus, the same HSF can sense the direction and attributes
of an EM emission from one user device (smartphone, laptop, IoT device),
and adaptively steer it toward an access point at the same time~\cite{CACM.2018,COMMAG.2018}.
This task is accomplished without adding field sensing hardware to
the setup~\cite{nanocom.2017}.

\section{Configuring HSFs for joint sensing and wave manipulation\label{sec:Proposed}}

\begin{figure}
\begin{centering}
\includegraphics[viewport=0bp 80bp 750bp 600bp,clip,width=1\columnwidth]{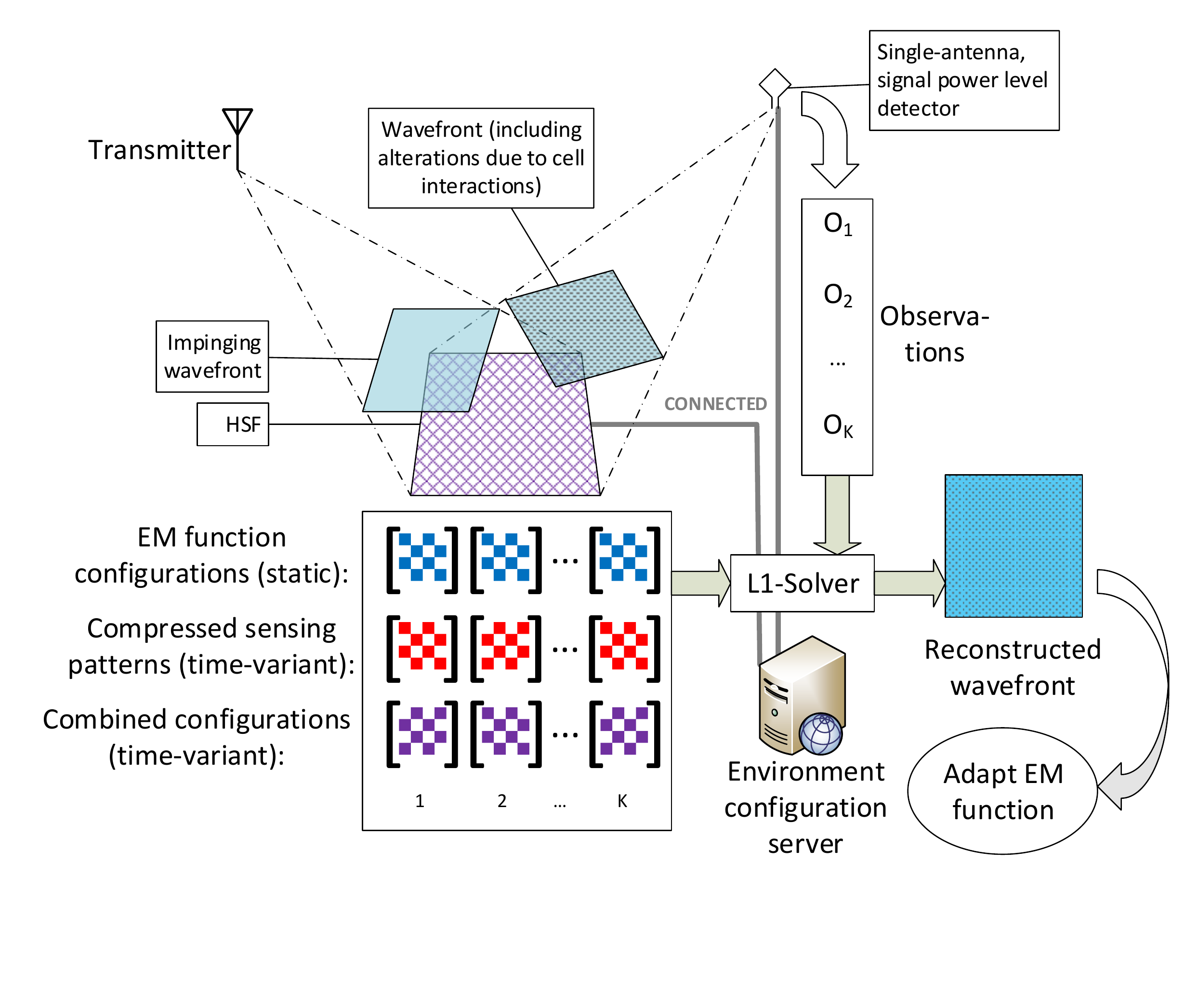}
\par\end{centering}
\caption{\label{fig:onePixelRF}The proposed joint sensing and wave manipulation
workflow via HyperSurfaces. }
\end{figure}
Consider a wireless communication setup illustrated in Fig.~\ref{fig:onePixelRF},
comprising a transmitter, a HSF and an environment configuration server.
The transmitter seeks to send data to a receiver (which is not shown
as it is irrelevant to the setup) while the HSF facilitates this communication
according to the programmable wireless environment concept shown in
Fig.~\ref{fig:OverviewPWE}. This can be accomplished by an EM function,
e.g., focusing and steering the EM wave impinging on the HSF towards
the receiver.

We will denote the HSF configuration that matches the intended EM
function as $C_{F}$. As discussed in Section~\ref{sec:prereqs},
$C_{F}$ is a $n\times n$ array, where each element $c_{F}^{ij}$
describes the state of the actuation element inside the HSF cell with
indexes $i,j$. Since HSFs with binary (ON/OFF) actuation elements
are more tractable to manufacture at low cost and large scales~\cite{yang2016programmable},
we will focus on configurations with binary elements (i.e., $1$ or
$0$). Under normal operation, i.e., without sensing duties, the environment
configuration server simply programs the HSF to follow the $C_{F}$,
and the required EM function materializes in the environment.

Let us now consider a scenario where the HSF performs wave sensing
only, following the RF imaging setup of Fig.~\ref{fig:onePixelRF}.
The objective is to detect the power of the impinging wave at each
of the HSF cells. Following the 1-pixel camera equivalent, the HSF
needs to undergo a series of configurations for sensing, denoted as
$C_{S}^{k},\,k=1\ldots K$. For each configuration, the server collects
the corresponding signal power measurements, $P_{1},P_{2},\ldots P_{K}$,
taken from an observing wireless device, denoted as \emph{detector}
in Fig.~\ref{fig:onePixelRF}. (Notice that the detector may coincide
with the transmitter or the receiver. However, we focus on the general
case where the detector is a separate user device). Finally, the server
reconstructs the impinging wavefront via the $L_{1}$ minimization
process described in Section~\ref{subsec:Compressive-Sensing}.

We proceed to study the following issues:
\begin{enumerate}
\item How can we transform the CS sampling matrix, $A$, (which is generated
by existing software~\cite{brad}) into the $C_{S}^{k},\,k=1\ldots K$
configurations for the HSF?
\item How can we combine $C_{F}$ with any $C_{S}^{k}$?
\end{enumerate}
For the first issue, notice that the elements of $A$ are real numbers
in general~\cite{brad}. Thus, we will seek to replace each row of
$A$ with a number of new rows with binary elements. We generalize
this process as a real-vector-to-binary-array decomposition process
and formalize it as Algorithm~\ref{alg:SLR}.
\begin{algorithm}[t]
\caption{\label{alg:SLR}The binary vector decomposition process{\footnotesize{}.}}

\hrule

\KwIn{A vector of reals, $\boldsymbol{v_{1\times N}}$; A number
of decimal digits to preserve, $I_{e}$; an acceptable error, $\epsilon$.}

\KwOut{A binary matrix, $B_{...\times N}$; a vector of reals $\boldsymbol{s}_{1\times...}$;
the real constants $S$,$D$,$U$.}

\hrule

$S\gets\min\left\{ \boldsymbol{v}\right\} $\;

$D\gets\max\left\{ \boldsymbol{v}-S\right\} $\;

\If{D=0}{$B\gets\emptyset_{1\times N}$\;$D\gets1$\;$U\gets1$\;$\boldsymbol{s}\gets1$\;$\boldsymbol{return}$\;}

$\boldsymbol{v}\gets round\left(\frac{\boldsymbol{v}}{D}\cdot10^{I_{e}}\right)$\;

$B\gets\emptyset$\;

$\boldsymbol{s}\gets\emptyset$\;

$\boldsymbol{l}^{prev}\gets\emptyset$\;

$coeff\gets1$\;

\While{$true$}{
$\boldsymbol{l}\gets\boldsymbol{v}:v_{i}>0$\;
\If{$\frac{\sum\boldsymbol{l}}{N}<\epsilon$}{$return$\;}
\If{$\boldsymbol{l}=\boldsymbol{l}^{prev}$}{$coeff\gets coeff+1$\;}
\Else{$\boldsymbol{l}^{prev}=\boldsymbol{l}$\;$B.add\_row\left(\boldsymbol{l}\right)$\;
$\boldsymbol{s}.add\_element\left(coeff\right)$\;
$coeff\gets1$\;}
$\boldsymbol{v}\gets\boldsymbol{v}-\boldsymbol{l}$\;
}

\hrule
\end{algorithm}
At lines 1-10, the original vector is normalized in the range $\left[0,1\right]$
(including a check for the case of a vector with all-equal elements
at line $3$). Subsequently, the normalized vector elements are scaled
by a power of $10$, depending on the number of decimal digits that
we wish to preserve during the decomposition. Over lines 15-30, the
process creates the rows of $B$ by marking the non-zero elements
of the scaled $\boldsymbol{v}$ with a binary '$1$' flag (line 16),
and promptly subtracting it from the scaled vector (line 29). The
process terminates if the number of non-zero elements in a new row
fall below a user-supplied amount, $\epsilon$ (line 17). Indicatively,
as shown later in Section~\ref{sec:Evaluation}, a value of $\epsilon=1\nicefrac{o}{oo}$
yields no discernible loss on the quality of the sensed outcome. Finally,
at lines 20-22 and 27, the process counts how many times a scaled
line appears (to avoid row repetitions in $B$), and returns it as
a vector $\boldsymbol{s}$. The original vector can be composed from
the process outputs as:
\begin{equation}
\boldsymbol{v}\approx\left(\sum_{\forall i}s_{i}\cdot\boldsymbol{B}_{\boldsymbol{row}\left(i\right)}\right)\cdot\frac{D}{U}+S\label{eq:decomp}
\end{equation}
Going back to equation (\ref{eq:CSmatrix}), assume that we treat
a row, $\boldsymbol{a}_{i}$, of $A$ as the vector $\boldsymbol{v}$
to be decomposed. Then, a single observation $o_{i}$ is equal to:
\begin{equation}
o_{i}=\boldsymbol{a}_{i}\cdot\boldsymbol{x}\stackrel{(\ref{eq:decomp})}{\approx}\left(\sum_{\forall i}s_{i}\cdot\boldsymbol{B}_{\boldsymbol{row}\left(i\right)}\cdot\boldsymbol{x}\right)\cdot\frac{D}{U}+S\cdot\left(\sum_{\forall m}x_{m}\right)\label{eq:itiscompl}
\end{equation}
where for future reference we denote the quantity:
\begin{equation}
X=\left(\sum_{\forall m}x_{m}\right)\label{eq:X}
\end{equation}
For the second issue, i.e., of combining $C_{F}$ and $C_{S}^{k}$
configurations, we follow an interleaving approach. First, we define
a \emph{mask,} $\mathcal{M}$, as a binary $n\times n$ array with
elements:
\begin{equation}
\mu_{ij}=\left\{ \begin{array}{c}
1,\,\text{iff \ensuremath{i} and \ensuremath{j}}\text{\,are even,}\\
0,\,\text{otherwise.}
\end{array}\right.\label{eq:mask}
\end{equation}
Subsequently, we redefine $C_{S}^{k}$ as a $m\times m$ matrix, where
$m=\nicefrac{n}{2}$ (assuming that $n$ is even, with no loss of
generality). In other words, we change our objective to sense the
impinging wave only at every other cell, given that this information
is still enough to characterize the impinging wave. Then, the combined
configuration for EM sensing and manipulation, $C_{S-F}$ is calculated
as:
\begin{equation}
C_{S-F}\left(C_{S}^{k},C_{F}\right):C_{F}\left(\mathcal{M}\right)\gets C_{S}^{k},
\end{equation}
i.e., the mask is treated as a 2D index for replacing every other
element of $C_{F}$ with those of $C_{S}^{k}$.

\begin{algorithm}[t]
\caption{\label{alg:Total}The joint wave sense-manipulate system operation{\footnotesize{}.}}

\hrule

\KwIn{The HSF side size, $n$ (in number of cells); A sampling matrix,
$A_{K\times\left(n^{2}\right)}$; An EM function configuration, $C_{F}$;
Decimal digits to preserve, $I_{e}$; Acceptable error, $\epsilon$.}

\KwOut{The sensed wavefront, $\mathcal{W}$.}

\hrule

$\mathcal{M}\gets equation\left(\ref{eq:mask}\right)$\;

$m\gets\nicefrac{n}{2}$\;

$\boldsymbol{o}\gets\emptyset_{1\times K}$\;

$X\gets ObtainMeasurement\left(\mathcal{M}\right)$\;

\For{$i=1\ldots K$}{$\left\{ B,\boldsymbol{\text{s}},S,D,U\right\} \gets BinaryDecomposition\left(\boldsymbol{a_{i}},I_{e},\epsilon\right)$\;$M\gets0$\;\For{$each\,row\,\boldsymbol{b_{i}}\,of\,B$}{$C_{s}\gets reshape\left(\boldsymbol{b_{i}},m\times m\right)$\; $P\gets ObtainMeasurement\left(\mathcal{C}_{S-F}\left(\mathcal{C}_{S},\mathcal{C}_{F}\right)\right)$\;$M\gets M+P\cdot s_{i}$\;$o_{i}\gets M\cdot\nicefrac{D}{U}+S\cdot X$\;}}

$\boldsymbol{x_{e}}\gets SparseReconstruct\left(\boldsymbol{o},A\right)$\;

$\mathcal{W}\gets reshape\left(\boldsymbol{\boldsymbol{x_{e}}},m\times m\right)$\;

\hrule
\end{algorithm}
Next, we proceed to formulate the complete system operation as Algorithm~\ref{alg:Total}.
At lines 1-3, the process initializes the mask, sets the vector of
observations to an empty state, and initializes $m$. At line 4, we
perform a special measurement once, to gain an estimate of the quantity
$X$ of equation (\ref{eq:X}). Since $X$ is essentially the sum
of all elements of the impinging wave, we perform a measurement with
the mask acting as the HSF configuration (i.e., all bits at the cells
to be sensed set to ON). At line 6, the process decomposes each row
of $A$ via Algorithm~\ref{alg:SLR}. Each binary decomposition is
reshaped as a $m\times m$ matrix (line 9), gets combined with the
$C_{F}$ and deployed to the HSF. A power measurement is obtained,
and the observation row is updated per element at line 12, following
relation (\ref{eq:itiscompl}). Finally, at line 15-16 the wavefront
is reconstructed and reshaped as a $m\times m$ matrix. (Notice that
reshaping pertains to splitting a vector into $m$ parts and concatenating
them vertically).

\section{Evaluation\label{sec:Evaluation}}

\begin{figure}
\begin{centering}
\includegraphics[width=1\columnwidth]{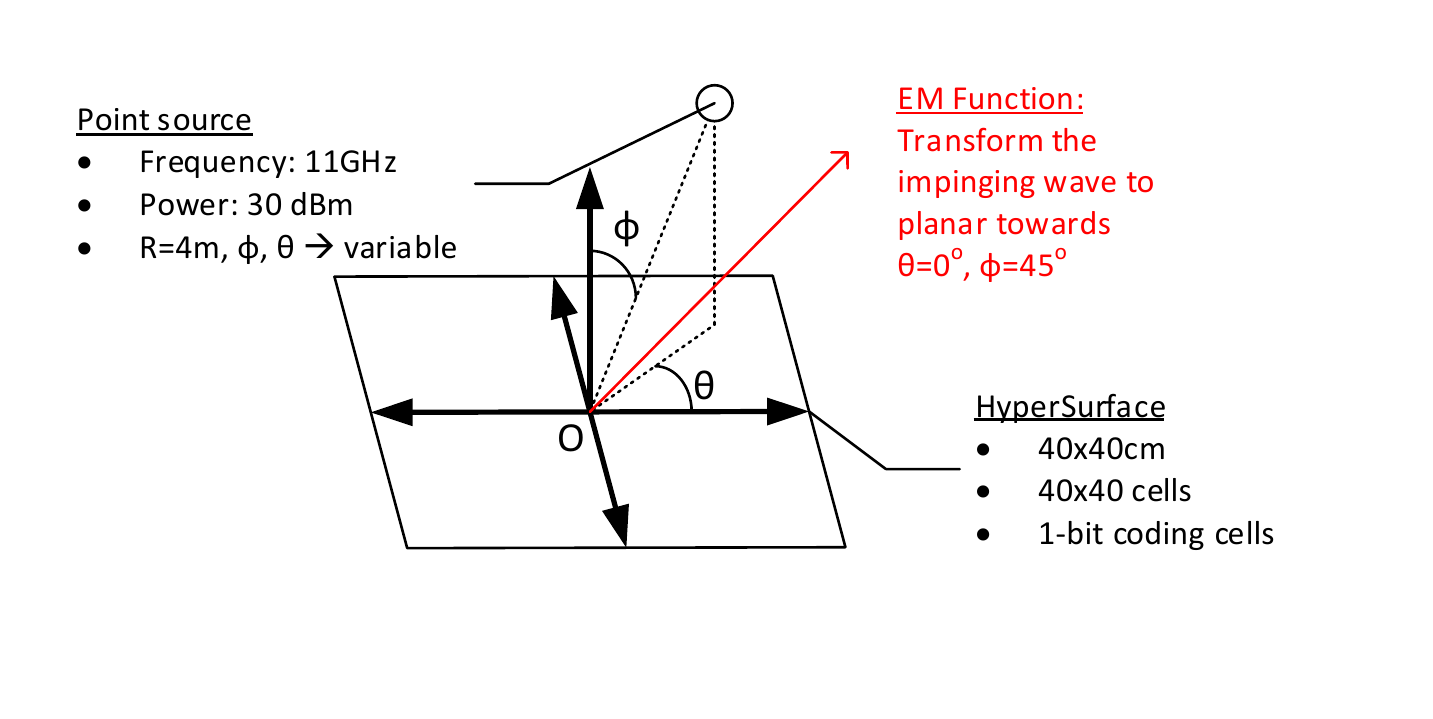}
\par\end{centering}
\caption{\label{fig:setup}Overview of the evaluation setup. }
\end{figure}
We proceed to evaluate the proposed scheme using simulations implemented
in MATLAB~\cite{MATLAB:2015}.

\subsection*{Setup}

We consider the setup of Fig.~\ref{fig:setup} comprising a HSF with
dimensions $40\times40\,\text{cm}$ and a number of cells equal to
$n^{2}=1600$ (i.e, regular $n\times n$ cell layout). Each cell contains
one active element (PIN diode) that can be either ON or OFF. Thus,
a HSF configuration is a $40\times40$ array of binary elements. This
setup replicates the one proposed by Li et al. in~\cite{yang2016programmable}.
This study offers an analytical way of:
\begin{itemize}
\item Calculating the configuration that matches a required EM function.
\item Calculating the scattering diagram of the HSF, for any input configuration.
\end{itemize}
One additional reason for picking this setup is that the aforementioned
analytical models have been validated via real EM measurements~\cite{yang2016programmable}.

We consider one point source (isotropic) whose electromagnetic attributes
are given in Fig.~\ref{fig:setup}. The detector is fixed at $\phi=0^{o}$,
$\theta=0^{o}$, $R=4\,\text{m}$. During the experiments, the source
will be placed at different points over a sphere with radius $R=4m$
centered at the HSF origin, $O$. Our objective is to reconstruct
the HSF-impinging wavefront via compressed sensing, and study its
relation to the position of the point source each time.

In all subsequent experiments, the EM function performed by the HSF
remains static. The HSF seeks to transform the point source emissions
to a planar wave departing towards the direction $\phi=45^{o}$, $\theta=0^{o}$.
The mask is also fixed to $m^{2}=400$ pixels (regular $m\times m$
grid). Furthermore, we set $K=300$, $I_{e}=2$ and $\epsilon=10^{-3}$.
The software of ~\cite{brad} is used for creating the sampling matrix
and performing the sparse signal reconstruction.

\begin{figure}
\begin{centering}
\subfloat[\label{fig:f20t0}$\phi:20^{o}$,$\theta:0^{o}$]{\begin{centering}
\includegraphics[width=2cm,height=2cm]{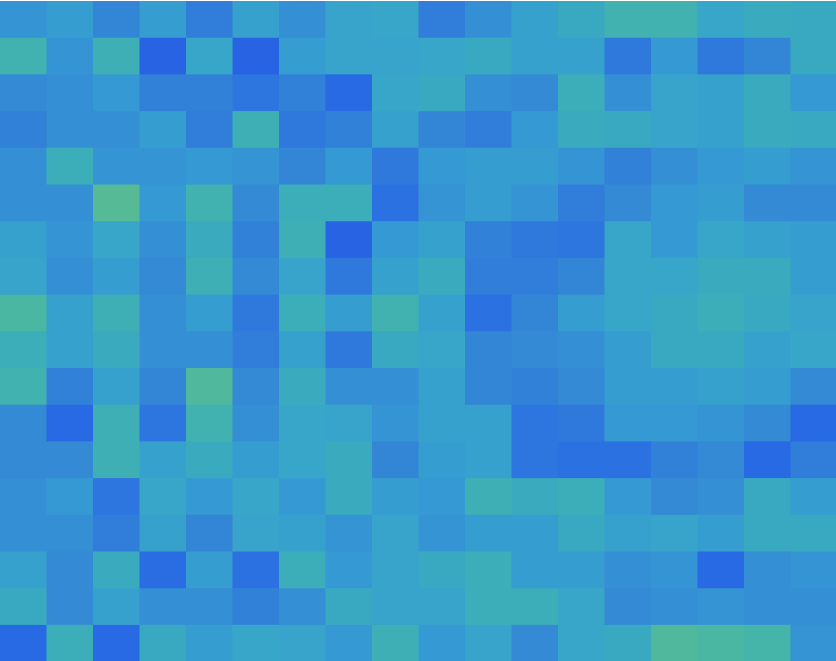}
\par\end{centering}

}~~~~~~\subfloat[\label{fig:f20t45}$\phi:20^{o}$,$\theta:45^{o}$]{\begin{centering}
\includegraphics[width=2cm,height=2cm]{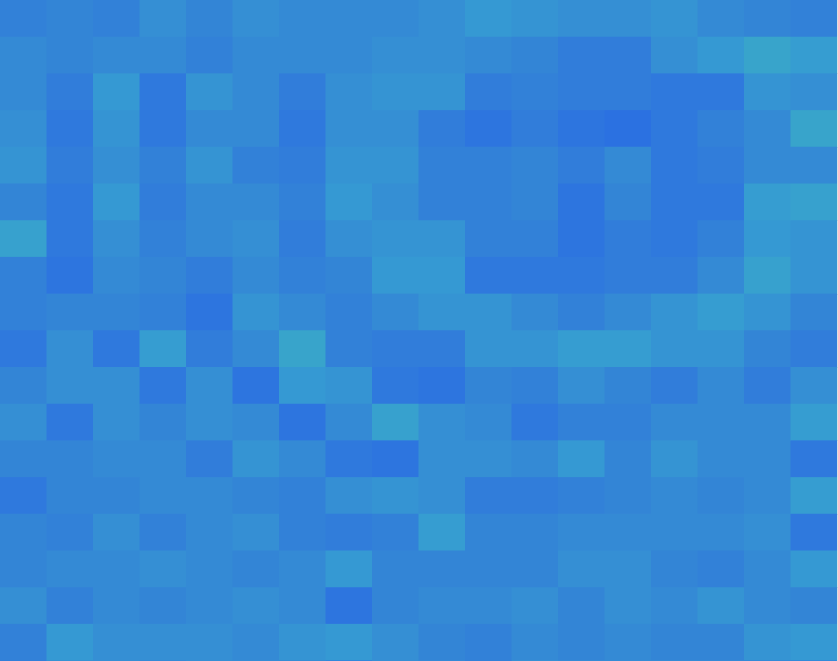}
\par\end{centering}
}~~~~~~\subfloat[\label{fig:f20t90}$\phi:20^{o}$,$\theta:90^{o}$]{\begin{centering}
\includegraphics[width=2cm,height=2cm]{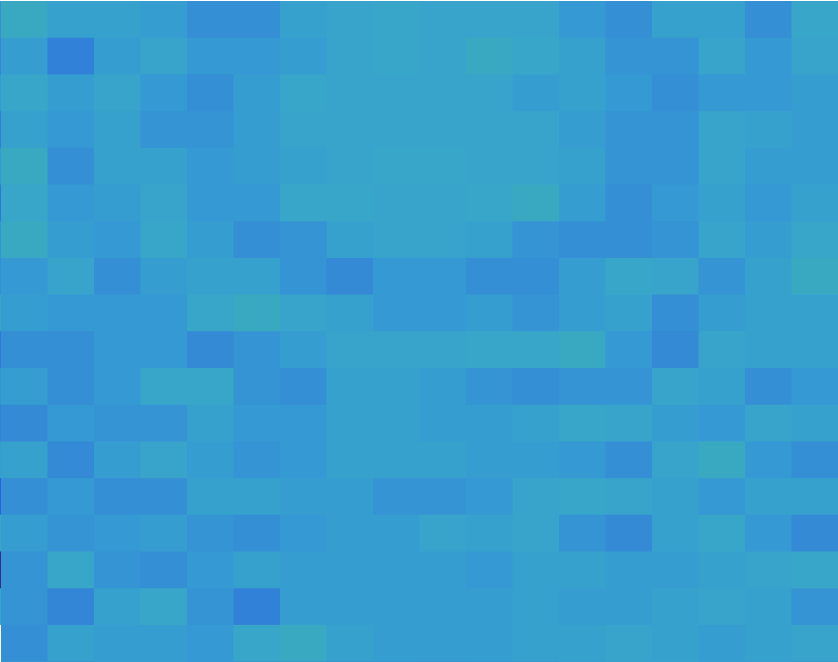}
\par\end{centering}
}
\par\end{centering}
\begin{centering}
\subfloat[\label{fig:f40t0}$\phi:40^{o}$,$\theta:0^{o}$]{\begin{centering}
\includegraphics[width=2cm,height=2cm]{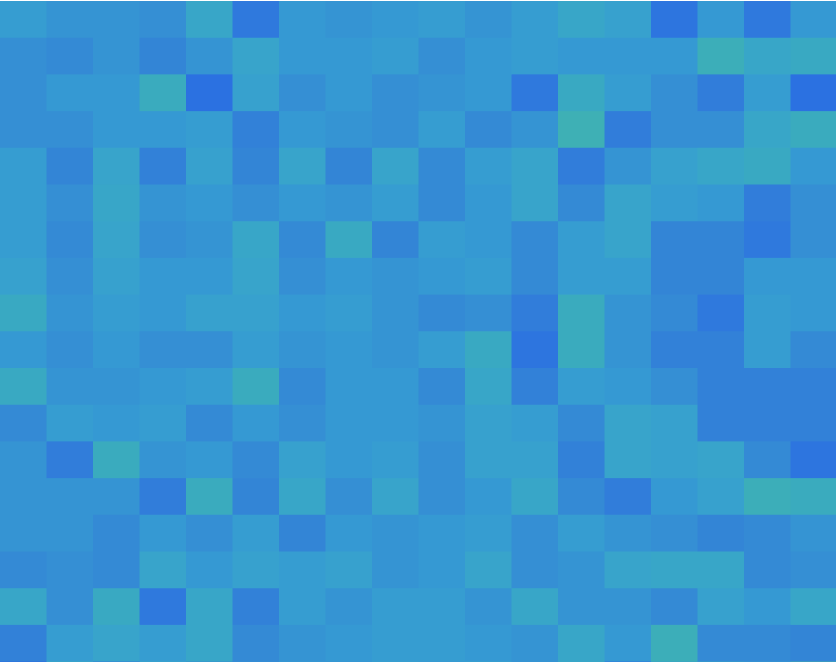}
\par\end{centering}
}~~~~~~\subfloat[\label{fig:f40t45}$\phi:40^{o}$,$\theta:45^{o}$]{\begin{centering}
\includegraphics[width=2cm,height=2cm]{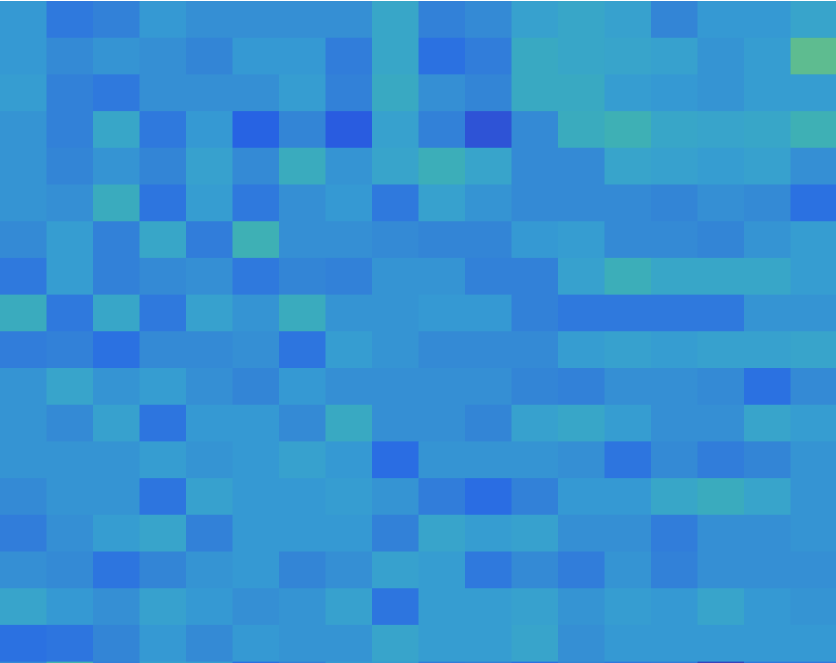}
\par\end{centering}
}~~~~~~\subfloat[\label{fig:f40t90}$\phi:40^{o}$,$\theta:90^{o}$]{\begin{centering}
\includegraphics[width=2cm,height=2cm]{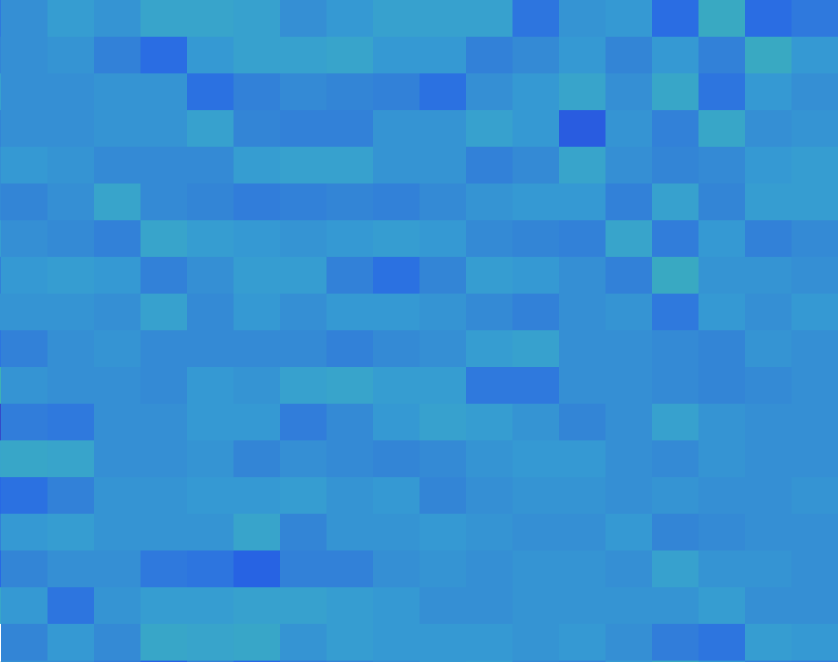}
\par\end{centering}
}
\par\end{centering}
\begin{centering}
\subfloat[\label{fig:f60t0}$\phi:60^{o}$,$\theta:0^{o}$]{\begin{centering}
\includegraphics[width=2cm,height=2cm]{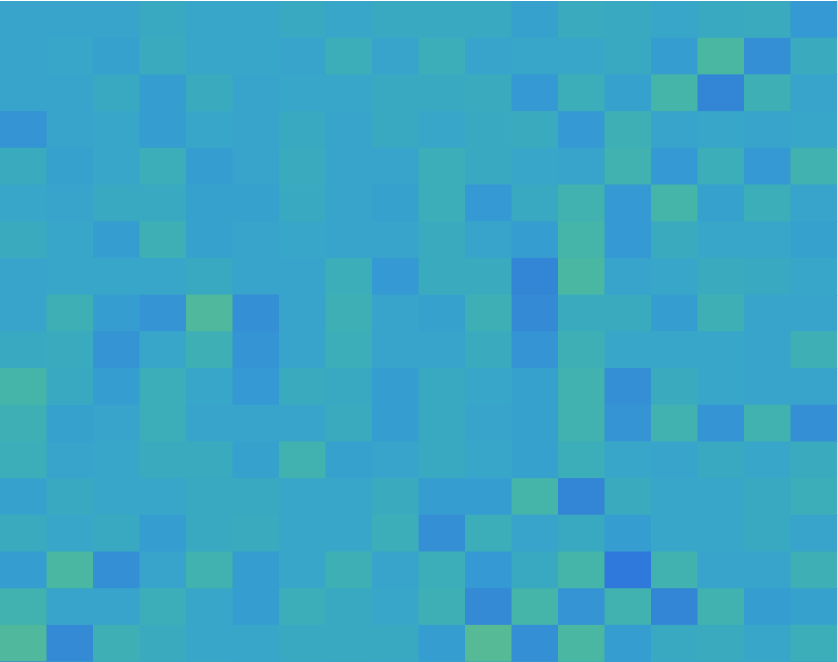}
\par\end{centering}
}~~~~~~\subfloat[\label{fig:f60t45}$\phi:60^{o}$,$\theta:45^{o}$]{\begin{centering}
\includegraphics[width=2cm,height=2cm]{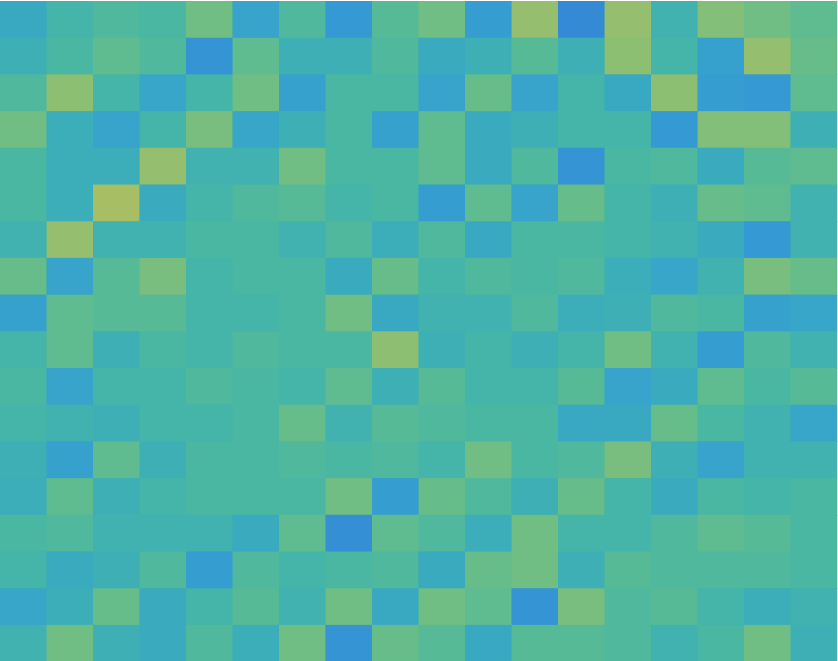}
\par\end{centering}
}~~~~~~\subfloat[\label{fig:f0t0}$\phi:0^{o}$,$\theta:0^{o}$]{\begin{centering}
\includegraphics[width=2cm,height=2cm]{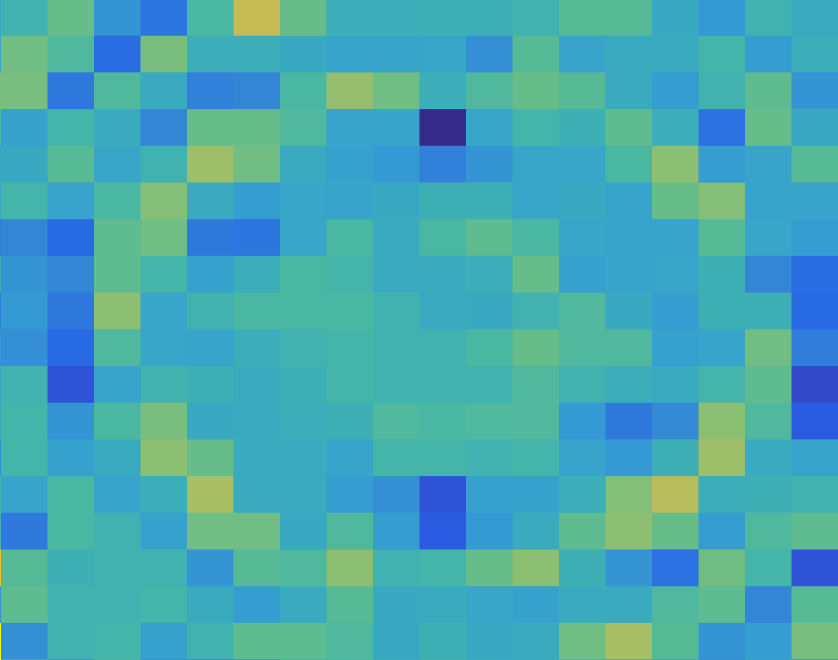}
\par\end{centering}
}
\par\end{centering}
\caption{\label{fig:Sensed-wavefronts}Sensed wavefronts reconstructed via
the proposed approach, for various locations ($\phi$,$\theta$) of
the point source. The distance of the point source from the origin
is fixed to $4\,\text{m}$.}

\end{figure}

\subsection*{Results}

Figure~\ref{fig:Sensed-wavefronts} shows the reconstructed wavefronts,
for various point source locations. When the source is located directly
above the HSF (Fig.~\ref{fig:f0t0}), the reconstructed wavefront
exhibits an expected circular form, as well as additional, unexpected
concentric circles. This effect is due to the interactions among the
HSF cells, discussed in the context of Fig~\ref{fig:MetaSurfCurrents},
which introduce artifacts into the sensed wavefront. Notice that the
HSF configuration seeks to transform the incident\textendash spherical\textendash wavefront
into a planar one, resulting into complex cross-cell interactions.

As the $\phi$ of the point source position increases (Fig.~\ref{fig:f20t0},
\ref{fig:f40t0}, \ref{fig:f60t0}), the concentric pattern is progressively
shifted towards the right, following the position of the source. Likewise,
when the $\theta$ of the point source location increases (Fig.~\ref{fig:f20t0},
\ref{fig:f20t45}, \ref{fig:f20t90} and \ref{fig:f40t0}, \ref{fig:f40t45},
\ref{fig:f40t90} and \ref{fig:f60t0}, \ref{fig:f60t45}), the concentric
pattern rotates counter-clockwise, again following the location of
the point source.

\begin{table}
\centering{}\caption{\label{tab:Effects}Effects of wave sensing on the EM manipulation
function efficiency.}
\begin{tabular}{|c|c|c|c|c|}
\hline
$\phi$ & $\theta$ & EM function efficiency & $\frac{n^{2}-m^{2}}{n^{2}}$ & $\sigma$ during sensing\tabularnewline
\hline
\hline
$20^{o}$ & $0^{o}$ & 75\% & 75\% & 0.00493\%\tabularnewline
\hline
$20^{o}$ & $45^{o}$ & 77\% & 75\% & 0.00477\%\tabularnewline
\hline
$20^{o}$ & $90^{o}$ & 73\% & 75\% & 0.00470\%\tabularnewline
\hline
$40^{o}$ & $0^{o}$ & 71\% & 75\% & 0.00517\%\tabularnewline
\hline
$40^{o}$ & $45^{o}$ & 70\% & 75\% & 0.00493\%\tabularnewline
\hline
$40^{o}$ & $90^{o}$ & 70\% & 75\% & 0.00454\%\tabularnewline
\hline
$60^{o}$ & $0^{o}$ & 70\% & 75\% & 0.00453\%\tabularnewline
\hline
$60^{o}$ & $45^{o}$ & 67\% & 75\% & 0.00487\%\tabularnewline
\hline
$0^{o}$ & $0^{o}$ & 89\% & 75\% & 0.00609\%\tabularnewline
\hline
\end{tabular}
\end{table}
\begin{figure}
\begin{centering}
\subfloat[\label{fig:PatNoSense}Scattering diagram of the HSF for no sensing. ]{\begin{centering}
\includegraphics[viewport=0bp 300bp 600bp 500bp,clip,width=1\columnwidth]{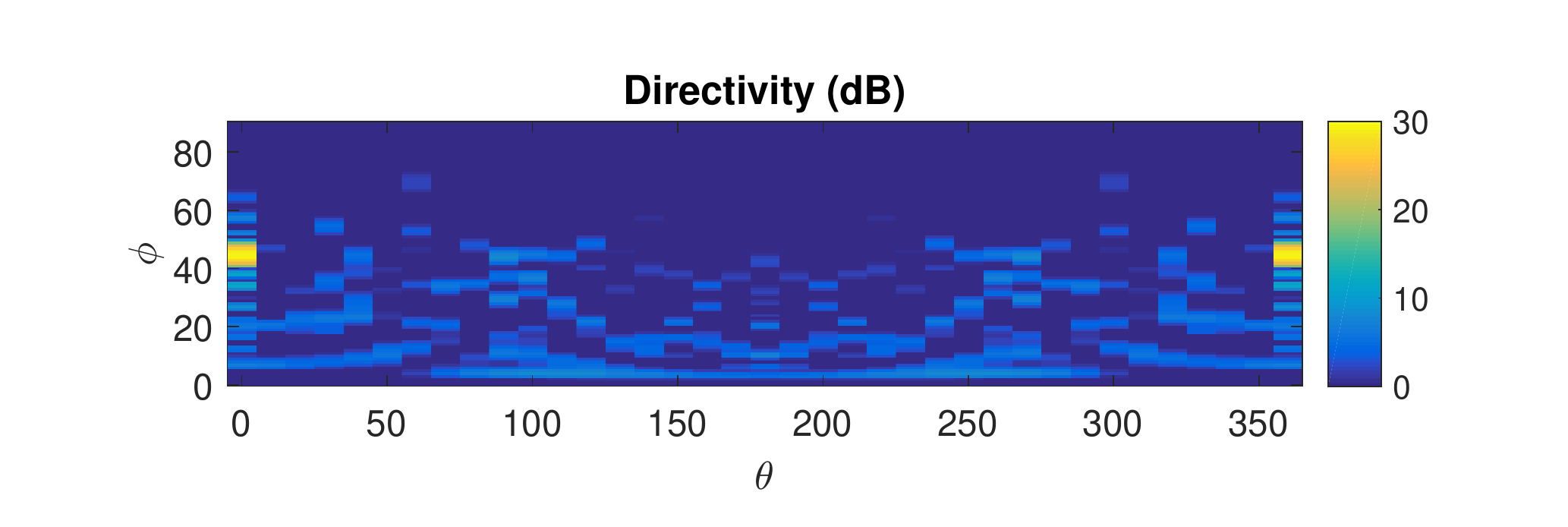}
\par\end{centering}
}
\par\end{centering}
\begin{centering}
\subfloat[\label{fig:PatSense}The same scattering diagram with sensing enabled. ]{\begin{centering}
\includegraphics[viewport=0bp 300bp 600bp 500bp,clip,width=1\columnwidth]{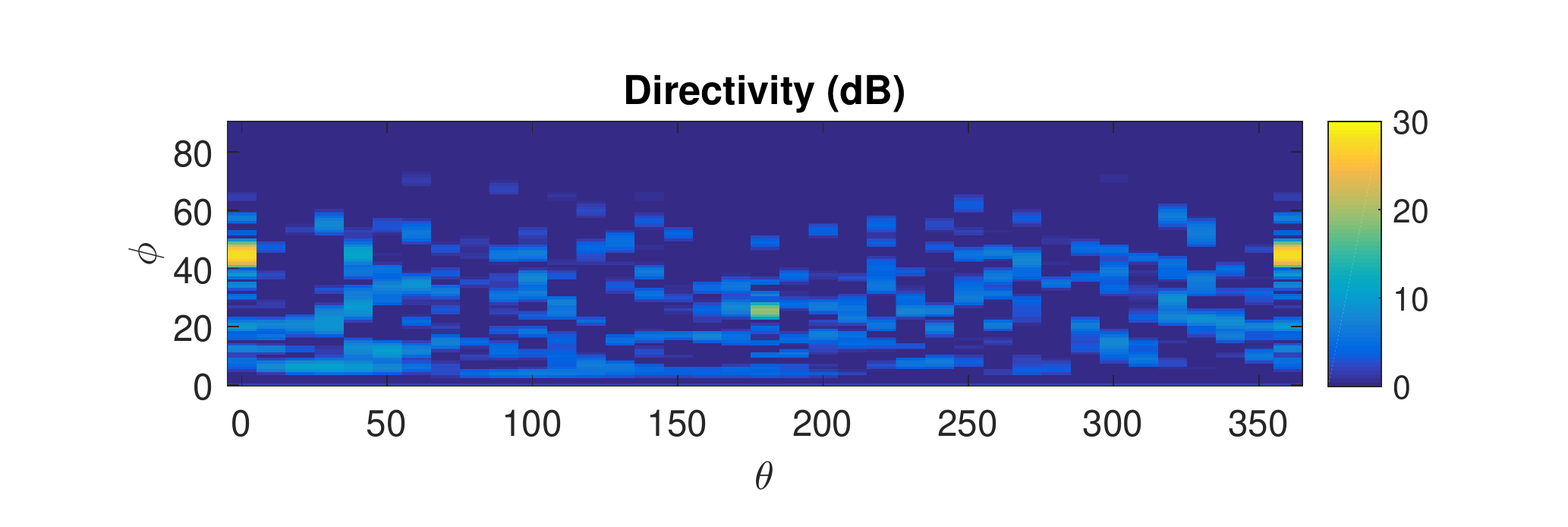}
\par\end{centering}
}
\par\end{centering}
\caption{\label{fig:Pstterns}Indicative effects of sensing on the scattering
pattern of the HSF for a given EM function ($\phi:0$, $\theta:0$). }
\end{figure}
We proceed to study the effects of combining wave sensing with the
EM function to the efficiency of the latter. To this end, we execute
one simulation with pure EM functionality without combined sensing
and we log the power, $P_{max}$, reflected towards the intended direction
($\phi:45^{o}$,$\theta:0$). Then, for each of the cases showed in
Fig.~\ref{fig:Sensed-wavefronts} (which include sensing), we execute
the same run and log the power reflected towards the same direction,
$P$. The ratio $\nicefrac{P}{P_{max}}$ is the EM function efficiency
column in Table~\ref{tab:Effects}. Notably, the efficiency is very
close to the ratio $\nicefrac{n^{2}-m^{2}}{n^{2}}$, which represents
the configuration array elements assigned to the EM function, after
subtracting the ones assigned to sensing. The EM function efficiency
when combined with sensing is thus approximately $75$\%. Variations
around this value are owed to the fact that sensing and wave manipulation
configurations can be occasionally aligned to each other (leading
to higher efficiency) or incompatible (leading to lower efficiency).
Additionally, we log the standard deviation of the EM function efficiency
over all sampling observations, and conclude that it is negligible
as shown in Table~\ref{tab:Effects}. In other words, the effects
of sensing on an EM function are static with regard to the sensing
configurations employed.

Finally, in Fig.~\ref{fig:Pstterns} we proceed to study the effects
of sensing on the whole HSF scattering diagram for one indicative
case (point source at $\phi:0$, $\theta:0$). Figure~\ref{fig:PatNoSense}
shows the scattering pattern when no sensing is employed, which naturally
contains parasitic lobes across which we have approximately $20\,\text{dB}$
lower power than the main lobe facing towards the intended direction
($\phi:45^{o}$,~$\theta:0$). The effects of introducing sensing
are shown in Fig.~\ref{fig:PatSense}. While the addition of sensing
duties introduces some additional parasitic lobes, they remain much
below the power of the main lobe, with a $15-20\,\text{dB}$ difference.

\subsection*{Discussion and future directions}

Programmable wireless environments promise full manipulation of EM
propagation in the form of wave \textquotedbl routing\textquotedbl ,
i.e., hopping from tile to tile while performing beam focusing~\cite{CACM.2018,COMMAG.2018,wowmom.2018}:
i) to maximize the power transferred to a remote device, either for
tele-charging or better quality of communication, ii) to minimize
the power received by a set of unintended devices, either for eavesdropping
mitigation or for interference cancellation. Additional applications
include: i) the mitigation of Doppler effects by ensuring that EM
waves are received from a direction perpendicular to the movement
trajectory of the receiver, and ii) EM wave scrambling and unscrambling
via destructive and constructive phase modification for advanced security
during the propagation via unsafe spaces~\cite{Liaskos2019ADHOC}.

In the present work, we studied a crucial requirement for enabling
these applications, i.e., the sensing of the EM wavefront that impinges
upon a tile. The proposed joint wave sensing and wave manipulation
was shown to extract useful information that strongly correlates with
the position of the point source. Moreover, the effects of the sense/manipulate
combination on the intended EM function can be manageable, as they
do not alter the HSF scattering pattern significantly. The combination
introduces a drop in efficiency of the EM function that is dictated
by the number of HSF cells assigned to sensing duties. This efficiency
drop can thus be minimized in HSFs with many cells in total, while
keeping the number of sensing cells constant.

In practical terms, the presented scheme can be deployed in two ways.
First, a receiving device can also act as the detector of the compressed
sensing process. In this case, the receiver can obtains incoming power
measurements and send them to the environment configuration server
for processing. The second way employs dedicated detectors placed
at few fixed points within the environment. Using beam-forming, they
can target tiles, obtained power measurements and send them to the
server for processing.

In the future, the sensed wavefronts will be processed via machine
learning techniques, to filter out the effects of the cell cross-interactions
and yield accurate wavefront measurements. Such measurements can then
be used for adaptively fine-tuning the HSF functions, to the benefit
of wireless devices. Moreover, the source sensing outcomes of the
proposed approach can be combined with external sensing systems to
improve their overall efficiency.

\section{Conclusion\label{sec:Conclusion}}

Programmable metasurfaces are the enablers of the Internet of Material
Properties, by introducing smart planar materials\textendash the HyperSurfaces\textendash that
can interact with impinging electromagnetic waves in a software defined,
adaptive manner. Highly promising applications include the programmable
wireless environments, wherein the electromagnetic propagation is
programmable, allowing for wave routing and introducing novel capabilities
in communication performance, security and power transfer. Central
to this paradigm is the sensing of waves that impinge upon a HyperSurface,
which typically rely on sensory hardware either internal or external
to the surface. The present work introduces a wave sensing that does
not employ hardware. Instead, it programs the HyperSurface in a manner
that interleaves a required electromagnetic behavior with a compressed
sensing workflow. The work showed that the impinging wave can be reconstructed
by simple power measurements at a single observation point, without
significantly affecting the desired electromagnetic behavior of the
HyperSurface.

\section*{Acknowledgment}

This work was funded by the European Union via the Horizon 2020: Future
Emerging Topics - Research and Innovation Action call (FETOPEN-RIA),
grant EU736876, project VISORSURF (http://www.visorsurf.eu).



\begin{thebibliography}{10}
\providecommand{\url}[1]{#1}
\csname url@samestyle\endcsname
\providecommand{\newblock}{\relax}
\providecommand{\bibinfo}[2]{#2}
\providecommand{\BIBentrySTDinterwordspacing}{\spaceskip=0pt\relax}
\providecommand{\BIBentryALTinterwordstretchfactor}{4}
\providecommand{\BIBentryALTinterwordspacing}{\spaceskip=\fontdimen2\font plus
\BIBentryALTinterwordstretchfactor\fontdimen3\font minus
  \fontdimen4\font\relax}
\providecommand{\BIBforeignlanguage}[2]{{%
\expandafter\ifx\csname l@#1\endcsname\relax
\typeout{** WARNING: IEEEtran.bst: No hyphenation pattern has been}%
\typeout{** loaded for the language `#1'. Using the pattern for}%
\typeout{** the default language instead.}%
\else
\language=\csname l@#1\endcsname
\fi
#2}}
\providecommand{\BIBdecl}{\relax}
\BIBdecl

\bibitem{akyildiz20165g}
I.~F. Akyildiz, S.~Nie, S.-C. Lin, and M.~Chandrasekaran, ``5g roadmap: 10 key
  enabling technologies,'' \emph{Computer Networks}, vol. 106, pp. 17--48,
  2016.

\bibitem{CACM.2018}
C.~Liaskos, A.~Tsioliaridou, A.~Pitsillides, S.~Ioannidis, and I.~F. Akyildiz,
  ``{Using any Surface to Realize a New Paradigm for Wireless
  Communications},'' \emph{{Communications of the ACM}}, vol.~61, pp. 30--33,
  2018.

\bibitem{COMMAG.2018}
------, ``{A New Wireless Communication Paradigm through Software-controlled
  Metasurfaces},'' \emph{{IEEE Communication Magazine}}, vol.~56, no.~9, pp.
  162--169, 2018.

\bibitem{wowmom.2018}
------, ``{Realizing Wireless Communication through Software-defined
  HyperSurface Environments},'' in \emph{{WoWMoM 2018, June 12-15, Chania,
  Crete, Greece}}, pp. 1--10.

\bibitem{tsilipakos2018pairing}
O.~Tsilipakos, A.~C. Tasolamprou, T.~Koschny, M.~Kafesaki, E.~N. Economou, and
  C.~M. Soukoulis, ``Pairing toroidal and magnetic dipole resonances in
  elliptic dielectric rod metasurfaces for reconfigurable wavefront
  manipulation in reflection,'' \emph{Advanced Optical Materials}, vol.~6,
  no.~22, p. 1800633, 2018.

\bibitem{Liaskos2019ADHOC}
C.~Liaskos, N.~Shuai, A.~Tsioliaridou, A.~Pitsillides, S.~Ioannidis, and
  I.~Akyildiz, ``A novel communication paradigm for high capacity and security
  via programmable indoor wireless environments in next generation wireless
  systems,'' \emph{Ad Hoc Networks}, vol.~87, pp. 1--16, may 2019.

\bibitem{shi2018accurate}
S.~Shi, S.~Sigg, L.~Chen, and Y.~Ji, ``Accurate location tracking from
  csi-based passive device-free probabilistic fingerprinting,'' \emph{IEEE
  Transactions on Vehicular Technology}, vol.~67, no.~6, pp. 5217--5230, 2018.

\bibitem{nanocom.2017}
A.~Tsioliaridou, C.~Liaskos, A.~Pitsillides, and S.~Ioannidis, ``{A Novel
  Protocol for Network-controlled Metasurfaces},'' in \emph{{ACM NANOCOM'17}},
  pp. 3:1--3:6.

\bibitem{duarte2008single}
M.~F. Duarte, M.~A. Davenport, D.~Takhar, J.~N. Laska, T.~Sun, K.~F. Kelly, and
  R.~G. Baraniuk, ``Single-pixel imaging via compressive sampling,'' \emph{IEEE
  signal processing magazine}, vol.~25, no.~2, pp. 83--91, 2008.

\bibitem{wallace2010analysis}
H.~B. Wallace, ``Analysis of rf imaging applications at frequencies over 100
  ghz,'' \emph{Applied optics}, vol.~49, no.~19, pp. E38--E47, 2010.

\bibitem{MSSurveyAllFunctionsAndTypes}
A.~Li, S.~Singh, and D.~Sievenpiper, ``Metasurfaces and their applications,''
  \emph{Nanophotonics}, vol.~7, no.~6, pp. 989--1011, jun 2018.

\bibitem{LiaskosAPI}
C.~Liaskos, A.~Tsioliaridou \emph{et~al.}, ``Initial {UML} definition of the
  {H}yper{S}urface programming interface and virtual functions,''
  \emph{European Commission, H2020-FETOPEN-2016-2017, Project VISORSURF:
  Accepted Public Deliverable D2.1, 31-Dec-2017, [Online:]
  \url{http://www.visorsurf.eu/m/VISORSURF-D2.1.pdf}}, 2017.

\bibitem{LiaskosComp}
C.~Liaskos, A.~Pitilakis \emph{et~al.}, ``Initial {UML} definition of the
  {HyperSurface} compiler middle-ware,'' \emph{European Commission,
  H2020-FETOPEN-2016-2017, Project VISORSURF: Accepted Public Deliverable D2.2,
  31-Dec-2017, [Online:] \url{http://www.visorsurf.eu/m/VISORSURF-D2.2.pdf}},
  2017.

\bibitem{donoho2006compressed}
D.~L. Donoho \emph{et~al.}, ``Compressed sensing,'' \emph{IEEE Transactions on
  information theory}, vol.~52, no.~4, pp. 1289--1306, 2006.

\bibitem{candes2008introduction}
E.~J. Cand{\`e}s and M.~B. Wakin, ``An introduction to compressive sampling [a
  sensing/sampling paradigm that goes against the common knowledge in data
  acquisition],'' \emph{IEEE signal processing magazine}, vol.~25, no.~2, pp.
  21--30, 2008.

\bibitem{tsaig2006extensions}
Y.~Tsaig and D.~L. Donoho, ``Extensions of compressed sensing,'' \emph{Signal
  processing}, vol.~86, no.~3, pp. 549--571, 2006.

\bibitem{brad}
\BIBentryALTinterwordspacing
E.~Brad. [Online]. Available:
  \url{https://statweb.stanford.edu/~candes/l1magic/}
\BIBentrySTDinterwordspacing

\bibitem{chan2009spatial}
W.~L. Chan, H.-T. Chen, A.~J. Taylor, I.~Brener, M.~J. Cich, and D.~M.
  Mittleman, ``A spatial light modulator for terahertz beams,'' \emph{Applied
  Physics Letters}, vol.~94, no.~21, p. 213511, 2009.

\bibitem{watts2012metamaterial}
C.~M. Watts, X.~Liu, and W.~J. Padilla, ``Metamaterial electromagnetic wave
  absorbers,'' \emph{Advanced materials}, vol.~24, no.~23, pp. OP98--OP120,
  2012.

\bibitem{sensale2013terahertz}
B.~Sensale-Rodriguez, S.~Rafique, R.~Yan, M.~Zhu, V.~Protasenko, D.~Jena,
  L.~Liu, and H.~G. Xing, ``Terahertz imaging employing graphene modulator
  arrays,'' \emph{Optics express}, vol.~21, no.~2, pp. 2324--2330, 2013.

\bibitem{li2016transmission}
Y.~B. Li, L.~L. Li, B.~B. Xu, W.~Wu, R.~Y. Wu, X.~Wan, Q.~Cheng, and T.~J. Cui,
  ``Transmission-type 2-bit programmable metasurface for single-sensor and
  single-frequency microwave imaging,'' \emph{Scientific reports}, vol.~6, p.
  23731, 2016.

\bibitem{yang2016programmable}
H.~Yang, X.~Cao, F.~Yang, J.~Gao, S.~Xu, M.~Li, X.~Chen, Y.~Zhao, Y.~Zheng, and
  S.~Li, ``A programmable metasurface with dynamic polarization, scattering and
  focusing control,'' \emph{Scientific reports}, vol.~6, p. 35692, 2016.

\bibitem{MATLAB:2015}
MATLAB, \emph{version R2015}.\hskip 1em plus 0.5em minus 0.4em\relax Natick,
  Massachusetts: The MathWorks Inc., 2015.

\end{thebibliography}
\end{document}